# MODEL BASED SYSTEM ENGINEERING APPROACH OF A LIGHTWEIGHT EMBEDDED TCP/IP

M. Z. Rashed

Computer science Dept, Faculty of computer science, Mansoura University, Egypt

Magdi_12003@yahoo.com

Ahmed E. Hassan

Electrical engineering Dept, Faculty of engineering, Mansoura University, Egypt

dr.hassan@ahmed-hassan.org

Ahmed I. Sharaf

Computer science Dept, Faculty of computer science, Mansoura University, Egypt

Ahmed.sharaf.84@gmail.com

## ABSTRACT:

*The use of embedded software is growing very rapidly. Accessing the internet is a necessary service which has large range of applications in many fields. The Internet is based on TCP/IP which is a very important stack. Although TCP/IP is very important there is not a software engineering model describing it. The common method in modeling and describing TCP/IP is RFCs which is not sufficient for software engineer and developers. Therefore there is a need for software engineering approach to help engineers and developers to customize their own web based applications for embedded systems.*

*This research presents a model based system engineering approach of lightweight TCP/IP. The model contains the necessary phases for developing a lightweight TCP/IP for embedded systems. The proposed model is based on SysML as a model based system engineering language.*

## 1-INTRODUCTION:

Embedded systems are collection of both software and hardware components that are tightly coupled which make it hard or very difficult to upgrade or replace any of the system components [16] .Embedded systems are special purpose computing devices which deigned to perform dedicated functions [21]. These systems vary in size, scope of use and complexity. Also these systems reside nearly in many devices. Embedded systems are usually resource limited in terms of processing power, memory, and power consumption, thus embedded software should be designed to make the best use of limited resources. A networked embedded system is a collection of spatially and functionally distributed embedded nodes, which are interconnected by means of wired or wireless communication infrastructure and communication protocol. Networked embedded systems can offer new services and more complex functions by means of connecting many other systems together. Networked embedded systems are most common in use and can be found in industrial planets, Mobile embedded devices, Telecommunication and wireless sensor network [25].

Networked embedded systems are based on a communication protocol. The most common and standard protocol is Internet Protocol Suite (TCP/IP). TCP/IP [17], [27] is known as a standard communication protocol for hosts to connect the Internet. TCP/IP is a set of communications protocols used for the Internet and other similar networks. TCP/IP protocol stack and its layers is discussed in the next section.

**TCP/IP protocol stack**

The TCP/IP model consists of four layers. From lowest to highest, these are the Link Layer, the Internet Layer, the Transport Layer, and the Application Layer [10]. The TCP/IP stack is viewed as a set of layers from telecommunication point of view .Each layer solves a set of problems involving the transmission of data, and provides a well-defined service to the upper layer protocols based on using services from some lower layers. Upper layers are logically closer to the user and deal with more abstract data, relying on lower layer protocols to translate data into forms that can eventually be physically transmitted.

Each layer form the stack consists of many protocols, each protocol solve a set of problems and has its detailed requirements and specifications. The whole TCP/IP protocol stack and its detailed protocols are defined in several documents which are called request for comment (RFC) provided by the IETF. Although the RFC document can describe the protocol specifications in accurate technique, it has some fundamental limitations [9].

Requirements traceability is established and maintained in the document based approach by tracing requirements between the specifications at different levels of the specification hierarchy. Requirements management tools are used to parse requirements contained in the specification documents and captures them in a requirements database. The traceability between requirements and design is maintained by identifying the part of the system or subsystem that satisfies the requirement, and/or the verification procedures used to verify the requirement, and then reflecting this in the requirements database.

Also completeness, consistency, and relationships between requirements, design, engineering analysis, and test information are difficult to assess since this information is spread across several documents. This makes it difficult to understand a particular aspect of the system and to perform the necessary traceability and change impact assessments. This, in turn, leads to poor synchronization between system level requirements and design and lower-level hardware and software design. It also makes it difficult to maintain or reuse the system requirements and design information for an evolving or variant system design. Also, progress of the systems engineering effort is based on the documentation status that may not adequately reflect the system requirements and design quality.

These limitations can result in inefficiencies and potential quality issues that often show up during integration and testing, or worse, after the system is delivered to the customer. Another approach used in modeling is called model based approach which is discussed in the next section.

**Model based system engineering**

Model based approach has been standard practice in electrical and mechanical design and other disciplines for many years. Mechanical engineering transitioned from the drawing board to increasingly more sophisticated two-dimensional (2D) and then three-dimensional (3D) computer-aided design tools beginning in the 1980s. Electrical engineering transitioned from manual circuit design to automated schematic capture and circuit analysis in a similar time frame. Computer-aided software engineering became popular in the 1980s for using graphical models to represent software at abstraction levels above the programming language. The use of modeling for software development is becoming more widely adopted, particularly since the advent of the Unified Modeling Language in the 1990s.

Model-based systems engineering (MBSE) is the formalized application of modeling to support system requirements, design, analysis, verification, and validation activities beginning in the conceptual design phase and continuing throughout development and later life cycle phases. MBSE is intended to facilitate systems engineering activities that have traditionally been performed using the document-based approach and result in enhanced communications, specification and design precision, system design integration, and reuse of system artifacts. The output of the systems engineering activities is a coherent model of the system (i.e., system model), where the emphasis is placed on evolving and refining the model using model based methods and tools.

MBSE provides an opportunity to address many of the limitations of the document-based approach by providing a more rigorous means for capturing and integrating system requirements, design, analysis, and verification information, and facilitating the maintenance, assessment, and communication of this information across the system's life cycle. Some of the MBSE potential benefits include the following:

1. Enhanced communications.
    1.1. Shared understanding of the system across the development team and other stakeholders.

1.2. Ability to integrate views of the system from multiple perspectives.
2. Reduced development risk.
    2.1. More accurate cost estimates to develop the system.
    2.2. Ongoing requirements validation and design verification.
3. Improved quality.
    3.1. More complete, unambiguous, and verifiable requirements.
    3.2. More rigorous traceability between requirement, design, analysis and testing.
    3.3. Enhanced design integrity.
4. Increased productivity.
    4.1. Reuse of existing models to support design evolution.
    4.2. Reduced errors and time during integration and testing.

MBSE [6] can provide additional rigor in the specification and design process when implemented using appropriate methods and tools [7]. The OMG announced SysML as a system engineering modeling language which is discussed in the next section.

**SysML**

The OMG Systems Modeling Language (OMG SysML) is a general purpose graphical modeling language [26], [20]. SysML [23],[29] can be used for specifying, analyzing, designing and verifying systems that may include hardware, software, information, personnel, procedures and facilities. This modeling language can integrate with any other engineering analysis model through a graphical and semantic foundation for modeling system requirement, behavior, structure and parametric. SysML has the following advantages over UML:

1. SysML's semantics are more flexible and expressive. SysML reduces UML's software-centric restrictions and adds two new diagram types, requirement and parametric diagrams.
2. SysML is a smaller language that is easier to learn and apply. Since SysML removes many of UML's software-centric constructs, the overall language is smaller as measured both in diagram types and total constructs

Model based approach is used in case of wireless sensor network by the center of embedded systems [24].SysML considered to be a lead modeling language. Sanda Mandutianu also used MBSE and SysML in modeling space mission systems and applying to a JPL space mission in early stages of formulation [19].

Authors of this work presented a detailed model based system engineering approach for lightweight embedded TCP/IP. The model is based on SysML. Section 2 discusses the previous related work. Section 3 discusses the proposed model and it details. Section 4 discusses the Future work. Section 5 discusses the conclusion

## 2-Related work:

Many authors and software vendors developed various versions of communication protocols for standard workstations and embedded systems.

Developing communication protocol is a special type of software development, which is a critical, complex and real time software. There is a gap between the art of software engineering and the practice of developing communication protocols [14]. Communication protocols usually developed using C language because of its time performance and fast execution time, which make it a good choice for developing communication protocols and embedded systems [5]. Developing TCP/IP for embedded systems is an important research area, which is discussed by many authors and micro-controller vendors.

Atmel provided an AVR internet toolkit [4], which can be used for 8 bit embedded internet applications. This toolkit is suitable for the manufacture products only. Aittamaa et al proposed a modular TCP/IP stack for embedded systems with a tiny timber interface [3]. Simon work added Point to Point Protocol (PPP) for embedded TCP/IP stack. Also Jakobsson et al developed a TCP/IP stack for a real time embedded systems [13]. Ognyan Dimitrov et al developed their own lightweight TCP/IP for embedded systems using PIC18F67J60 [22].

One proposed approach to solve the problem of communication protocol developing is using a formal description language technique to create the protocol specification. SDL [8], Estell [12] and Lotos [11] are examples of formal description languages used for this problem. Since SDL is the most widely adopted language, there is no reusability [14] in the design phase and very limited in the implementation phase.

Adams Dunkels presented two proposal based on the first approach for this problem which called UIP [1] and IwIP [18]. His proposals are based on redesigning TCP/IP as lightweight separate software that is targeting tiny 8 bit microcontrollers. There are three weak points in Dunkels's proposals. First, there is no software model. Since communication protocols are complex software, modeling is necessary process. Software modeling and analysis can improve system maintenance, flexibility, extensibility and ease of system understanding. Second, both UIP and IwIP are targeting tiny embedded systems that make it difficult to difficult to use the solution for anther microcontroller architecture. Lastly, there is no modularity in which makes it hard to customize the functionality of system. It could be clearer if the authors of the system presented the architecture.

The work presented in this paper is a SysML model of a lightweight TCP/IP for embedded systems. The proposed top level model is discussed in section 3. The components details and its model are discussed in section 4.

## 3.) Proposed Top level model

This section presents an overview of proposed system. This section contains requirement, use case and package diagram of the system.

## 3.1) Requirement Diagram:

Requirement diagrams are a new feature added to SysML and not existing in UML2.0 [28], requirement diagrams can describes the requirement of the systems from many different viewpoints. The top level requirement diagram shows the system requirement, user requirement and developer requirement, each category has its specific requirement to

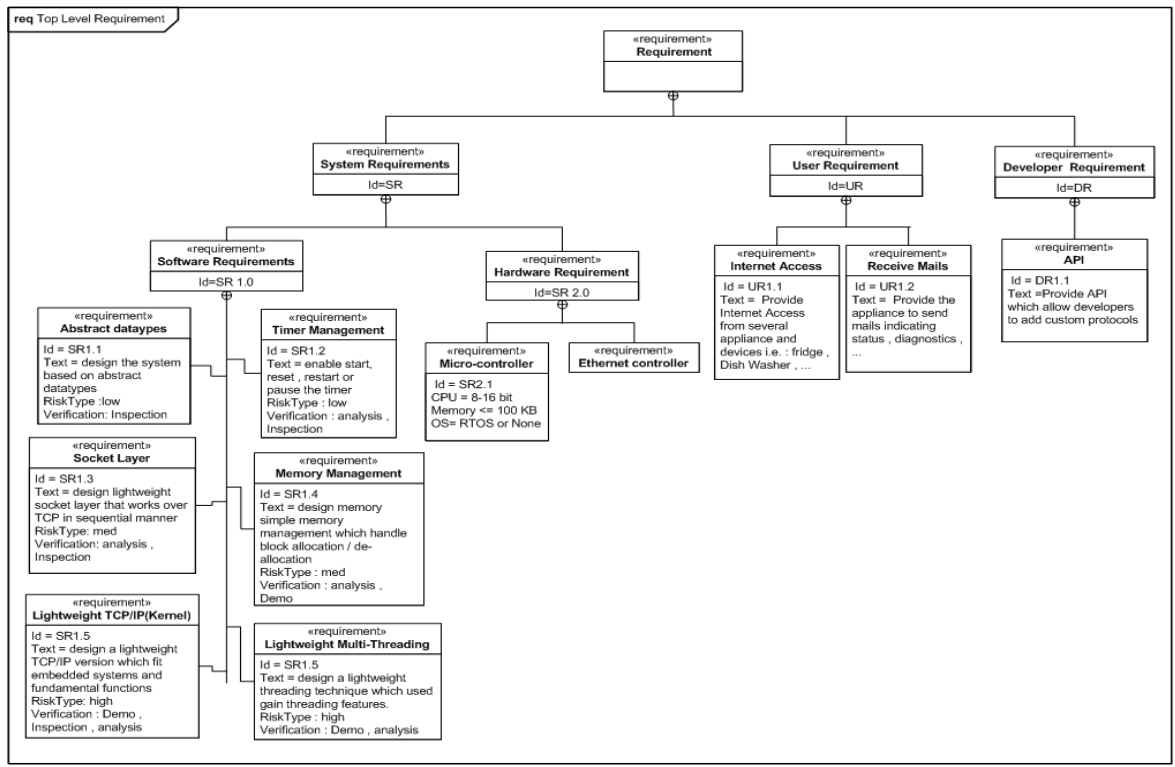

Figure1  Top level requirement diagram

gain the best usage from the system as shown in Figure 1.The requirement properties shown include some standard SysML requirement prosperities such as "Id","Text", "RiskType" and "Verification". The enumeration list of each requirement property is shown in Table 1.

| Requirement property | Enumeration list | Description |
|---|---|---|
| Id | None | The sequence Id of requirement item |
| Text | None | Short description of requirement item |
| RiskType | | The expected risk that could occurs by developer |
| | High | |
| | Med | |
| | Low | |
| Verification | | The verification method/technique used |
| | Analysis | Indicates that verification will be performed by verification of requirement and sub requirements under |

|  |  | specific conditions. |
|---|---|---|
|  | Demo | Indicates that verification will be performed in real environment and with the use of laboratory equipment. |
|  | Inspection | Indicates that verification will be performed by examination of the component, comparing the appropriate characteristics with a pre-determined standard. |
| CPU |  | The central processing power properties |
| OS |  | The type of the used operating system |

Table 1 list of requirement prosperities and its enumeration list

### 3.2) System behavior:

This section presents the overall system behavior though use case diagram. Use case is a behavior diagram which is common between UML2.0 and SysML. The top level use case of the proposed system is shown in Figure2.

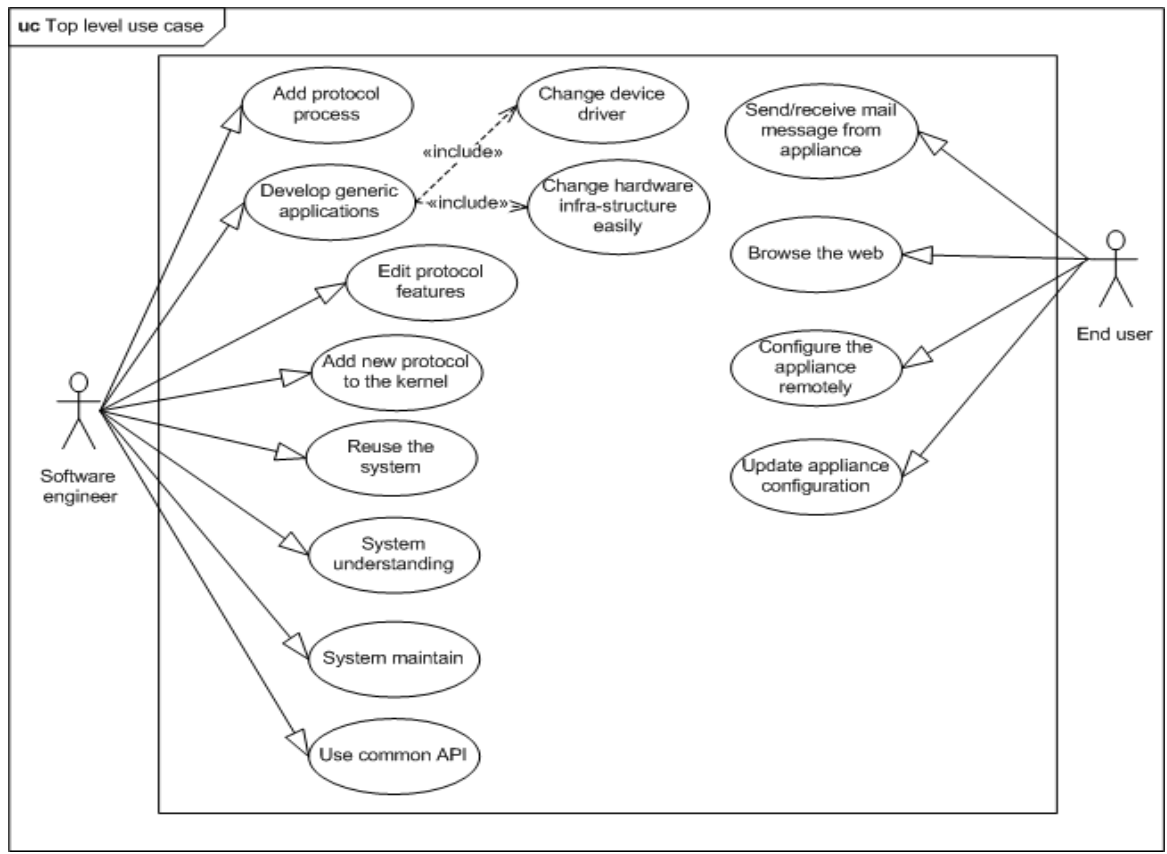

Figure2 Top level use case of the proposed system

### 3.3) System Structure:

This section presents the overall system structure through package diagram. Figure 3 shows the package diagram of the system structure. The package diagram includes the components of the system from software engineering point of view. The system contains five componenets which are:

- Hardware
- Device driver
- TCP/IP core layer which includes
  o Time manager
  o Memoery manager
  o Multi-threading
  o Lightweight TCP/IP kernel.
- Socket layer.
- User application layer.

The next section discusse each componenet of the system componenets with its details and model. User application componenet is out of this paper scope and is not discussed.

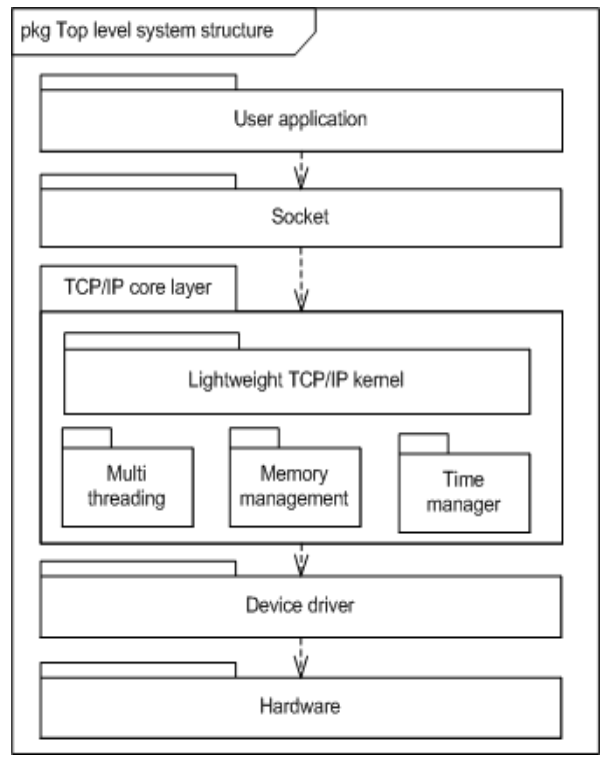

Figure3 Top level package diagram of the proposed system.

**4.) Components details:**

In this section the system components will be discussed in some manner of details. Each component will be introduced through structure and behavior models.

**4.1) Hardware**

The hardware layer includes the basic hardware components. The internal block diagram is shown in Figure 4. The used approach presents a simple I/O mapping using a simple microcontroller and Ethernet controller. For very fast throughput, as may be required for voice or video streaming, a small microprocessor will not be fast enough. This is because microprocessors require many instructions to transfer data from one location to another. Using a FPGA with an internal CPU core (an embedded processor) could perfrom high thoughput and fast accessing. Arriving frames are pre-processed by the FPGA hardwired logic (first processor), and stored in dual port RAM for further processing by the program based embedded core. Pipelined functions such as checksums, header address comparisons etc are handled by the first processor in real time. Operations such as management of state tables, connection establishment and other higher-layer software intensive functions are handled by the second processor. Designs like these can be fast enough for real-time 100 Mbps and even 1000 Mbps systems.

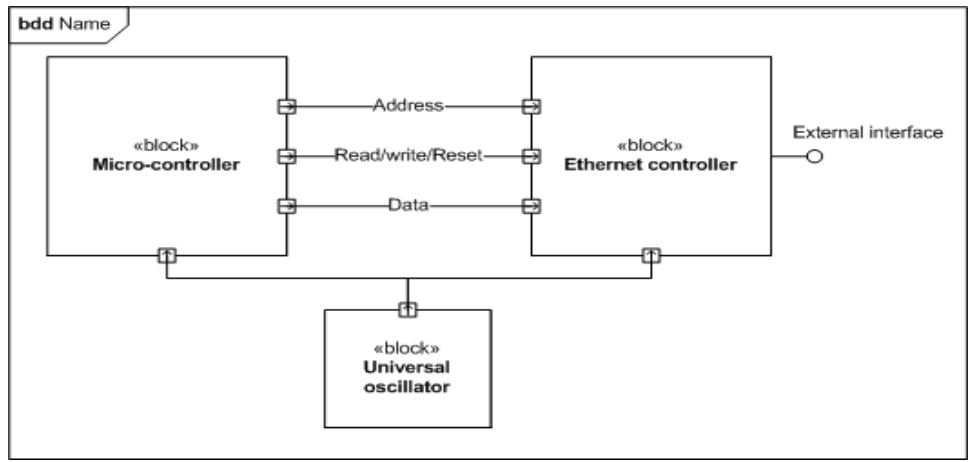

Figure 4 I/O mapping between a microcontroller and Ethernet controller

**4.2) Device driver**

This section describes techniques used for modeling the device driver behavior, use case is shown in Figure 5. The supported methods in this use case are discuused as follows:

- **Direct hardware accessing**

This category contains direct hardware accessing methods like chip hard reset, read from register and write to register.

- **Initializing and status**

This category contains initializing method and return status.

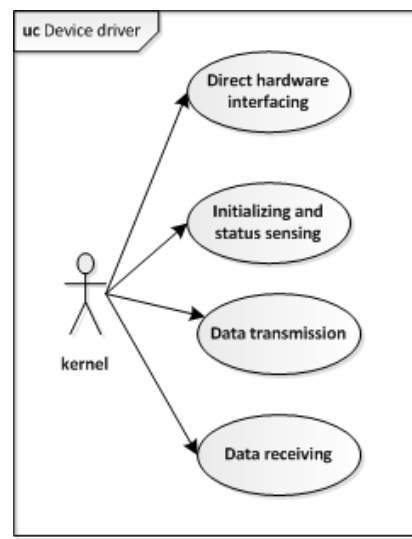

- **Data Transmission**

This category contains three basic methods which are initializing registers and memory, read from DMA and perform physical transfer.

- **Data receiving**

This category also contains three basic methods which are polling the device, initializing the registers and perform data transfer and memory de-allocation.

Figure 5 device driver use case

Another aspect of device driver behavior is activity diagram. Figure 6 shows activity diagram of sending frame, Figure 7 shows activity diagram of receiving frame.

**4.3) TCP/IP core layer**

This layer consists of lightweight TCP/IP (kernel), timer manager, multi-threading manager and memory manager. This section discusses the details of those layer components.

### 4.3.1) Lightweight TCP/IP (kernel)

This layer is considered as the most important component, which is discussed in some manner of details. The kernel component contains the data structures and functions used by TCP and IP protocols. Figure7 shows the IP use case. The IP methods which are use by the kernel are: "*IPH_Init*", "*IPH_Handler*", "*IPH_FillPacket*", "*IPH_FillHdr*" and "*IPH_ChkSum*". The description of these methods is shown in Table1 Appendix A. The relation between the kernel and IP can be illustrated from two scenarios which are sending and receiving data.

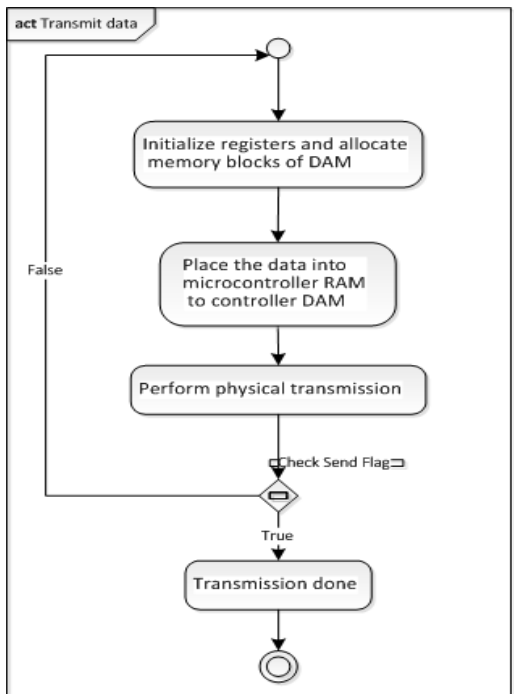

Figure 6 send data activity diagram

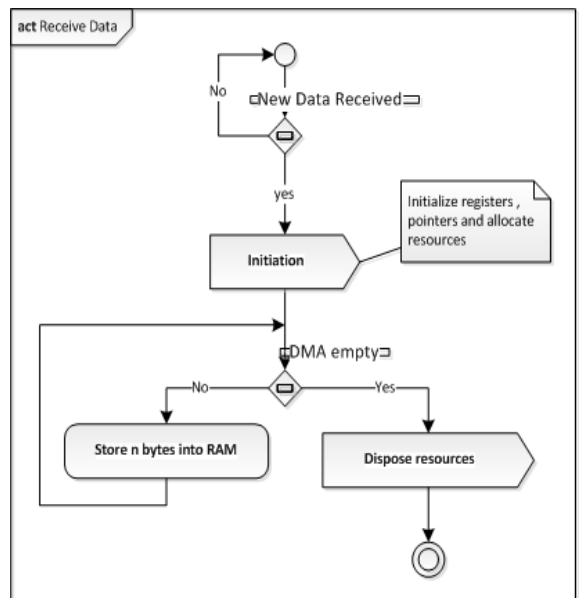

Figure 7 receive data activity diagram

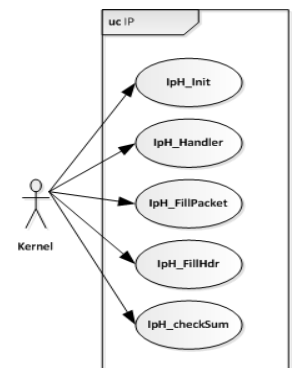

Figure 8 IP use case

First scenario, discuss how should the kernel send data from application program or user process. Figure 8 shows the detailed sequence diagram of sending data using IP. Second scenario, discuss how the kernel should handle the received IP packet and perform required processing. Figure 9 shows the activity diagram of receiving data. In this scenario, the kennel extracts some information from the received packet such source and destination IP address, protocol type and so on. The kernel then calls the corresponding module to handle the data

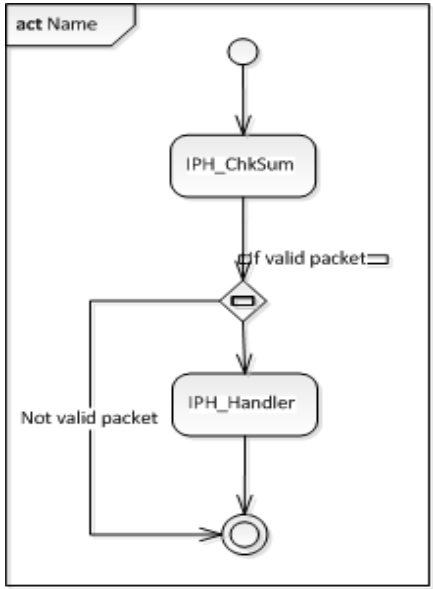

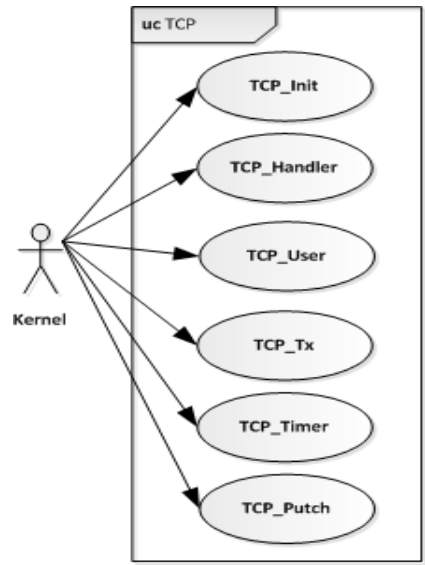

Figure 10 TCP use case diagram

Figure 9 activity diagram of receiving data

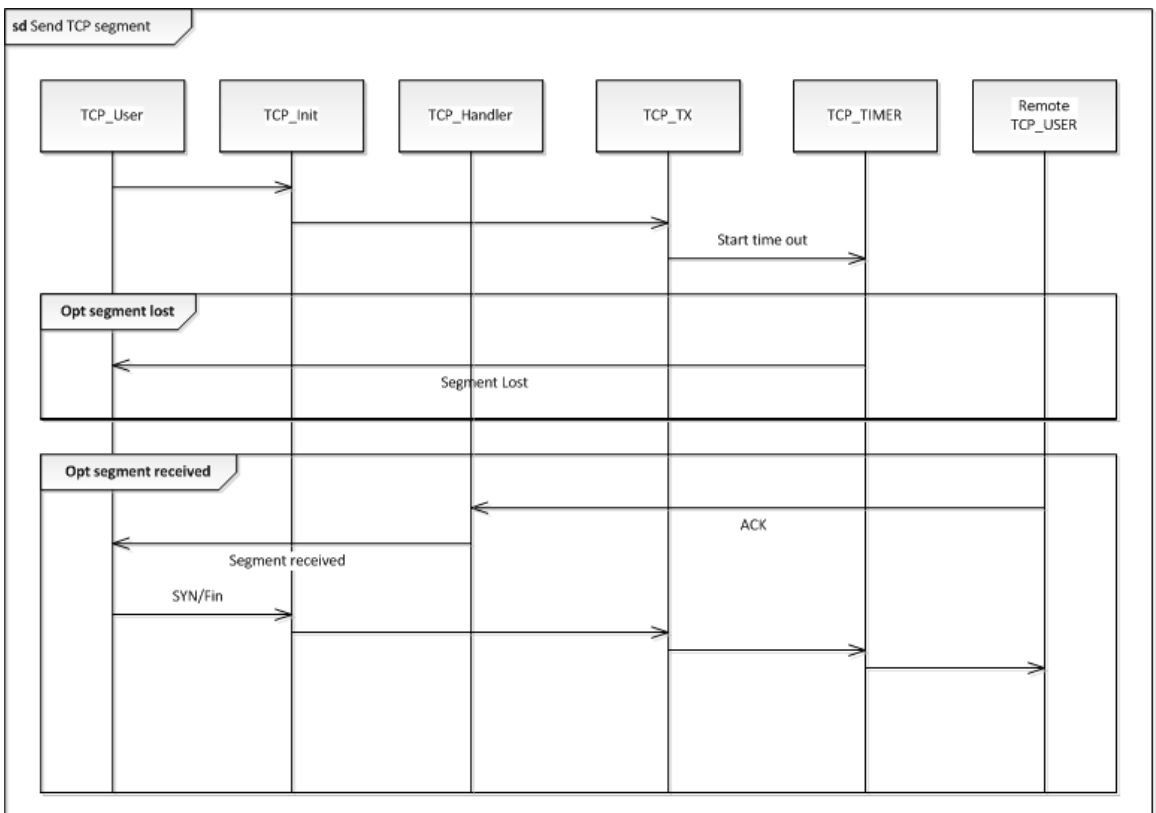

Figure 11 TCP Sequence diagram

TCP is a little different protocol than IP. TCP is a point-to-point, connection oriented, reliable, and byte stream protocol. The proposed use case of TCP is shown in Figure 10. This use case illustrates the interaction between the kernel and TCP. Table 2 Appendix A shows the description of this use case. The relation between the kernel and TCP is illustrated into send/receive scenarios. First scenario, discuss how should the kernel send TCP segment typically from TCP_User to remote TCP_User. Figure 11 illustrates the used sequence diagram to send TCP segment.

The second scenario, discuss how the kernel should handle a received TCP segment. Figure 12 illustrates the used approach.

**4.3.2) Multi-threading**

Multi-threading [15] technique is very useful approach used in general development. Embedded systems need collaborative task execution with especial features. Embedded systems requires more lightweight software that can be managed and executed in limited processing power such as 8-16 bit microcontroller. This work contains a lightweight multi-threading technique. The multi-threading state machine is shown in Figure12.

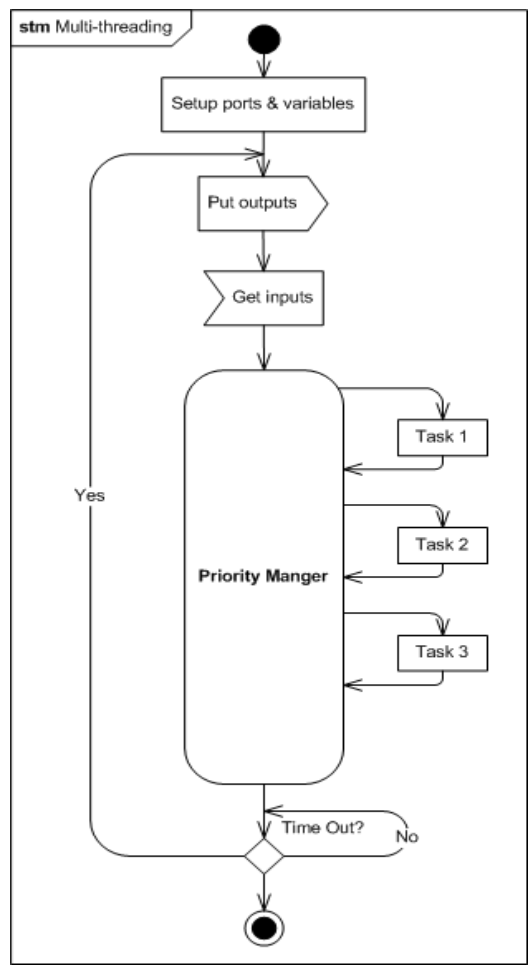

Figure 12 Multi-threading state machine diagram

This work contains a lightweight multi-threading technique which is based on Protothread [2] . The proposed multi-threading technique is based on duff device principal. Duff device state that a case statement is still legal within a sub-block of its matching switch statement. This technique is stack less thread which provides linear code execution for event driven systems. One advantage of these threads is there is no need to implement thread per stack as ordinary thread. In memory constrained systems, the overhead of allocating multiple stacks can consume large amounts of the available memory.

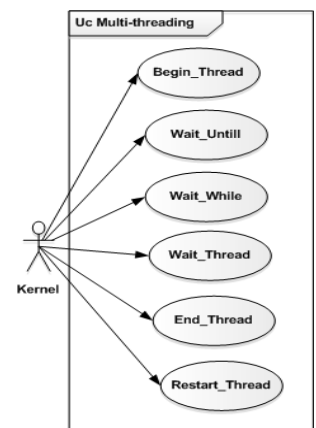

Figure 13 Multi-threading use case

The Protothread mechanism does not specify any specify method to invoke or schedule a Protothread this is defined by the system using Protothread. If a Protothread is run on top of an underlying event-driven system, the Protothread is scheduled whenever the event handler containing the Protothread is invoked by the event scheduler. The supported functions are shown in Figure 13 which illustrates the use case diagram of the proposed multi-threading technique. The internal block diagram of the multi-threading is shown in Figure 14.

**4.3.3) Timer manager**

One critical point in any communication protocol process is time management. Protocol process depends on time for many cases such as connection time out, resend data and wait for acknowledgement. Connection oriented protocol usually use time manager in order to control connection time, determine waiting time. Therefore time management component in TCP/IP is important. Calculating the intervals is determined by the hardware clock. Time manager should provide basic operations such as set time interval, restart the timer and reset the timer with the last configured interval value. These operating are shown in Figure 14 which represents the use case of Time manager module.

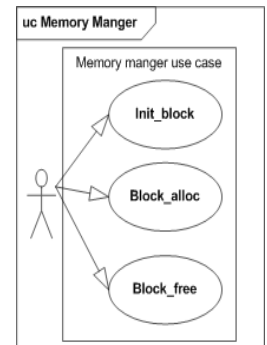

Figure 14 Timer manager use case

**4.3.4) Memory manager**

Memory is the most critical hardware component in any computer system. The kernel could access the memory in order to perform any of the following functions:
- Initialize memory block.
- Allocate block
- De-allocate block

Figure 15 show the use case of memory manager functions and how to deal with memory block, "Init_Block" which represents the declaration of memory block to be handled, "Block_alloc" which represents the allocation of memory that already declared and "Block_free" which represents the memory de-allocation of declared block. The proposed behavior for the memory manager is to use single buffer for holding packets , when a packet is arrived the device driver place it in the global buffer and call the kernel modules. If the packet contains data, the kernel notifies the corresponding application. When the application receives a notify message it have to take one action from the following:

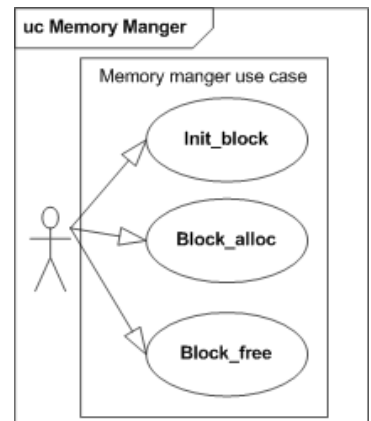

Figure 15 memory management use case use

- Perform online processing on global buffer
- Copy the packet contents to secondary buffer and perform the processing on it.

Figure 16 shows the proposed activity diagram which illustrates the workflow in memory manager.
One other view is also important in memory management is the functionality of the proposed architecture which demonstrate what will the module do with memory blocks.

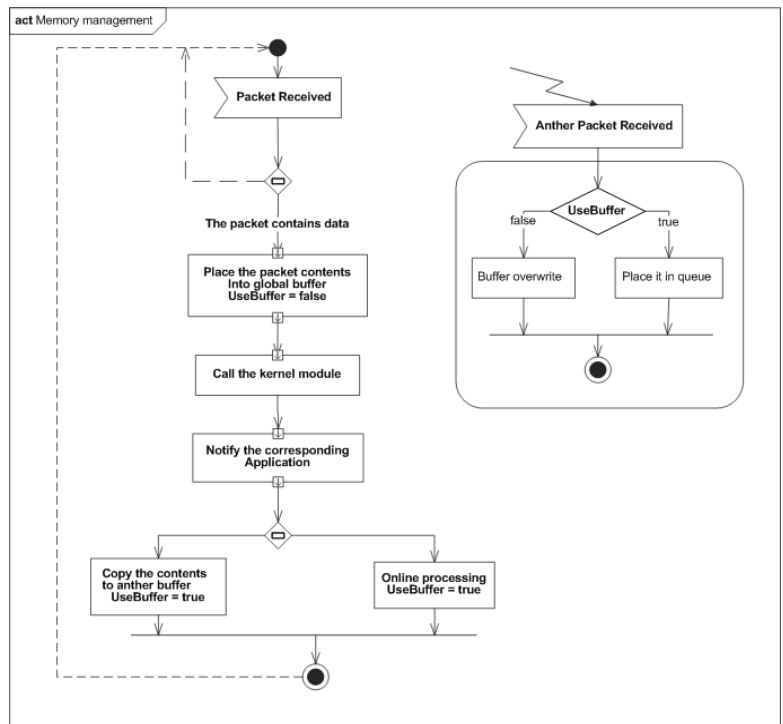

Figure 16 activity diagram of memory management.

### 4.4) Socket Layer

The next layer in the proposed architecture is called socket layer. This layer represents socket and the functions used to handles socket. Socket is defined as end point of a bidirectional communication link between hosts or between processes on the same host. Socket component is based heavily on the thread mechanism that make socket inherits the complexity of thread, the socket being by invoking "*Begin_thread*" and terminated by "*End_thread*". Figure 16 shows the internal block diagram of socket manager. Figure 17 shows the corresponding use case which illustrate the main functionality of socket management component.

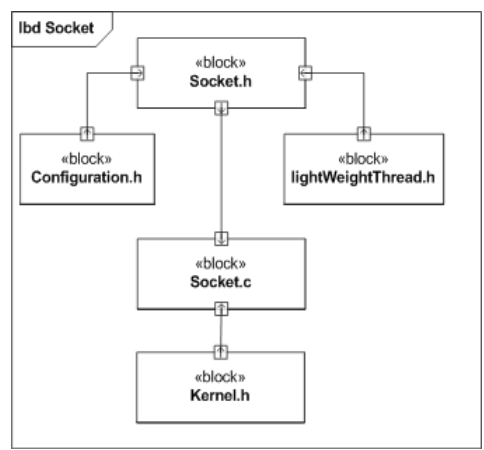

Figure 16 internal block diagram of socket manager

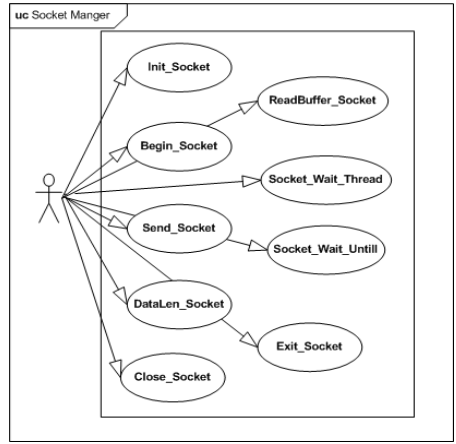
Figure17 socket management use case

# FUTURE WORK

We plan to develop this model for a real 8 bit microprocessor and complete the implementation phase. We also plan to compare our implementation result against any similar work from complexity point of view and run time performance.

# CONCLUSION

We presented in this research paper anther view of lightweight TCP/IP to fit embedded systems. Although TCP/IP protocol stack is described in several RFC documents, the text based approach has many limitations. The model based approach is used here using SysML as modeling language to model a lightweight TCP/IP architecture for embedded systems environment. The model presents a requirement, analysis and design phases of lightweight TCP/IP. Also the main components and protocols are modeled using SysML. The presented model illustrate SysML befits in modeling communication protocols and TCP/IP protocol stack.

# APPENDIX A

| Method Name | Description |
|---|---|
| *IPH_Init* | This method initialize the IP header |
| *IPH_Handler* | This method is called when IP packet is received, checks the protocol field and calls the corresponding module such ICMP, TCP … and so on. |
| *IPH_FillPacket* | This method is used to fill the packet data. |
| *IPH_FillHdr* | This method is used to fill the IP packet header |
| *IPH_ChkSum* | This method is used to validate the packet data and order by applying packet check sum |

Table 1 IP methods descriptions

| Method Name | Description |
|---|---|
| *TCP_Init* | This method initialize the TCP header |
| *TCP_Handler* | This method is used to handle the TCP segment, it also handle the TCP SYN, FIN, RST and normal ACK status. |
| *TCP_User* | This method is used to notify the TCP applications with the correct data. |
| *TCP_putch* | This method is used to send character over TCP |
| *TCP_Tx* | This method is used to send TCP segment |
| *TCP_Timer* | This method is used to handle timer timeout during SYN received and ACK wait status. |

Table 2 TCP methods description

# REFERENCES:

1. A.Dunkels. UIP a TCP/IP stack for 8-and16-bit microcontrollers. [Web page] http://www.sics.se/~adam/uip/index.php/Main_Page
2. Adam Dunkels et al, 2006. Protothreads: Simplifying Event-Driven Programming of Memory-Constrained Embedded Systems. Acm
3. Aittamaa, Simon and Rova, Isak. 2007. A modular TCP/IP stack for embedded systems with a tinyTimberinterface. Lulea University of Technology, Sweden.
4. Atmel. [Web page] http://www.atmel.com.
5. Barr. 2006. Programming Embedded Systems. O'Reilly.
6. Buede, Dennis M. 2009 The Engineering design of Systems models and methods. John Wiley & Sons.
7. E. Andrianarison , J-D. Piques, 2010, SysML for embedded automotive Systems: a practical approach. ERTS².
8. Ellsberger, Jan. 1997.SDL: Formal Object-Oriented Language for Communicating Systems. Prentice Hall PTR
9. Friedenthal, Sanford et al. 2010. A Practical Guide to SysML , Elsevier
10. Insam, Edward, 2003. TCP/IP Embedded Internet Applications. Elsevier.
11. ISO, 1989 Lotos. ISO international standard IS8807
12. ISO, 1989 Estell. ISO international standard IS8807.

13. Jakobsson, Stefan and Dahlberg, Erik. 2007. Development of a TCP/IP Stack in real time embedded system. Umea University, Department of Computing Science , Sweden
14. K.hramboulidis and A.Mikroyannidis. 2003. Using UML for the Design of Communication Protocols: The TCP Case Study. International Conference on Software, Telecommunications and Computer Networks.
15. Keith Curtis, 2006. Embedded Multitasking. Elsevier.
16. Li, Qing and Yao, Carolyn.2003. Real-Time Concepts for Embedded Systems, CMP Books.
17. Loshin, Pete.2003. TCP/IP Clearly Explained, Elsevier.
18. LwIP a lightweight TCP/IP stack. [Web page] http://www.sics.se/~adam/lwip/
19. Mandutianu, Sanda. 2009. Modeling Pilot for Early Design Space Missions. 7th Annual Conference on Systems Engineering Research.
20. Marcos V. Linhares et al 2007. Introducing the Modeling and Verification process in SysML. 12th IEEE Int. Conf. on Emerging Technologies and Factory Automation
21. Noergaard, Tammy et al. 2005. Embedded systems Architecture, Elsevier.
22. Ognyan Dimitrov et al 2008. Embedded Internet based system , Annual conference of the University of Rousse
23. Omg, 2005  Systems Modeling Language (SysML) Specification
24. Sachdeva et al 2005.System Modeling.  A Case Study on a Wireless Sensor Network. Center for Embedded Computer System University of California, Irvine.
25. Sridhar, T 2003. Designing Embedded Communications Software. CMP Books ,
26. SysML. [Web page] http://SysML.org.
27. Tanenbaum, Andrew S. and Wetherall, David J. 2010. Computer Networks (5th Edition) Prentice Hall.
28. Tim Weilkiens 2006. Systems Engineering with SysML/UML .Elsevier.
29. Vanderperren, Yves and Dehaene, Wim. 2005, SysML and Systems Engineering Applied to UML-Based SoC Design. Citeseer.